# The Early Years of Quantum Monte Carlo (1): the Ground State

*Michel Mareschal[1],[2],*

*Physics Department, ULB , Bruxelles, Belgium*

# Introduction

In this article we shall relate the history of the implementation of the quantum many-body problem on computers, and, more precisely, the usage of random numbers to that effect, known as the Quantum Monte Carlo method.

The probabilistic nature of quantum mechanics should have made it very natural to rely on the usage of (pseudo-) random numbers to solve problems in quantum mechanics, whenever an analytical solution is out of range. And indeed, very rapidly after the appearance of electronic machines in the late forties, several suggestions were made by the leading scientists of the time , - like Fermi, Von Neumann, Ulam, Feynman,..etc-  which would reduce the solution of the Schrödinger equation to a stochastic or statistical problem which , in turn, could be amenable to a direct modelling on a computer.

More than 70 years have now passed and it has been witnessed that, despite an enormous increase of the computing power available, quantum Monte Carlo has needed a long time and much technical progresses to succeed while numerical quantum dynamics mostly remains out of range at the present time. Using traditional methods for the implementation of quantum mechanics on computers has often proven inefficient, so that new algorithms needed to be developed. This is very much in contrast with what happened for classical systems. At the end of the fifties, the two main methods of classical molecular simulation, Monte Carlo and Molecular Dynamics, had been invented and an impressively rapid development was going to take place in the following years: this has been described in previous works [Mareschal,2018] [Battimelli,2018].

The simulation of quantum N-body systems is intrinsically more difficult than the simulation of classical models: particles are not localised and one has therefore to deal with wave functions.

---

[1] Retired

[2] email: mmaresch@ulb.ac.be

Physical properties have to be averaged over the wave function, which generally depend on *3N* variables. The corresponding Monte Carlo sampling has to be specifically adapted in order to be efficient. Besides, the symmetry properties of fermions and bosons imply different wave functions and different statistical properties leading to different physical behaviour. This requires additional sampling- sampling the exchange of particles- to be made whenever physics depends on those symmetry properties.

In this article, we will describe a few important steps in the history of quantum Monte Carlo. We will focus on works which have proven essential in solving issues which were also relevant from a physical point of view: the exact description of the electron gas at zero temperature is such an example. This result has been obtained after the extension to fermions of the Monte Carlo method which was first developed for bosons and which amounted to project any trial wave function to the ground state of the system.

The phase transition of Helium 4 from normal fluid to superfluidity at the so-called lambda temperature (2.2 K) is another example of a standing problem solved by simulation. Here the method is based on the path integral formulation of quantum mechanics by Richard Feynman. Working at a finite temperature, there is no more restriction to the ground state of the system. The Path Integral Monte Carlo (PIMC) method is technically different from the variational or diffusion Monte Carlo approaches. In order to ease the reading of this paper, we prefer to divide the exposition in two parts and delay the part on PIMC into a separate and second article referred to as QMC2.

David Ceperley was associated to those two successes. We base this article, as well as QMC2, on an interview of him, which we have made recently. We do not pretend to be exhaustive in our presentation of the history of quantum Monte Carlo: we wish to focus on a few works which have been decisive in the success of quantum simulations. Our view is therefore biased and we may be unfair to groups and scientists which have contributed quite significantly to Quantum Monte Carlo. In order to correct for this, we shall provide direct and indirect references of related work. Our article is in the line of previous articles published on various historical aspects of molecular simulations[Battimelli,2018; Levesque,2018; Mareschal, 2018]. Let us also mention a recent book which appeared on related topics [Battimelli, 2020].

The article is organised following the historical order. Section 2 is dedicated to the early times in the years 50s. After those historical preliminaries, we present the work of McMillan who combined Metropolis Monte Carlo methods and the variational principle to obtain the ground state of a model of Helium 4 atoms. An important step forward was then the Green Function Monte Carlo method which was developed a few years later by Malvin Kalos for an ensemble of bosons. The next part deals with the extension of this work to a fermion system: this was achieved in the thesis work of David Ceperley in the late 70s. This led to important results obtained on the electron gas and on the description of hydrogen. The methods used evolved to a technique very similar to Brownian dynamics: it is being referred to as Diffusion Monte Carlo and we provide a few details on its formulation. We end with a few remarks of historical content.

# From Fermi to Kac: early works

The first Monte Carlo simulation on a computer was performed at the initiative of John Von Neumann in the years 1947-48. It concerned the direct simulation of the fission reaction : many of the elementary steps, like, for example, the reactive events when a neutron collides with an Uranium nucleus, were modeled as stochastic processes and use was made of (pseudo-)random numbers to that effect.

Von Neumann's simulation proposal was described in a letter written to the director of the theory division in Los Alamos, Dr. Richtmyer. The letter has been since declassified[3]; it is written with so many details that one can figure out precisely what the computer is really doing. The simulation itself took several years to be performed, it involved several people, among whom John Von Neumann and his wife Klara, Nick Metropolis, Stan Ulam and a few others. It was performed on various computers, starting from the historic ENIAC, within a rapidly-evolving hardware architecture as well as programming technique: at some stage, it involved the creation of the stored-program concept [Haigh, 2014].This work is well documented and has been already analyzed by Peter Galison [Galison, 1996] as symptomatic of the paradigm shift induced by the computer in the physical sciences (the so-called *trading zone*).

Those years of post-war period were years of great intellectual freedom and creativity. It also led to the organisation of several scientific meetings where engineers, chemists, physicists and mathematicians would discuss the possibilities offered by the new electronic machines[Hurd,1985]. Among topics being discussed, the use of statistical sampling and probability theory to produce solutions to partial differential equations and integro-differential equations[4].

Stan Ulam published in 1949 the first article with the name Monte Carlo in it : « *The Monte Carlo Method*» [Metropolis,1949], written with Nick Metropolis, describes in a qualitative way the possibilities offered by the use of random numbers to solve mathematical and physical problems. The core of the article is concerned with the case of physical processes which are stochastic in essence, like the diffusion of neutrons in media where they are subjected to random influences: the probability model is then inspired from the process itself. However, the article suggests also a more sophisticated approach, namely the possibility to invent a probability model whose solution would be related to the solution of a complex partial differential equation. This is clearly the case for quantum mechanics and it becomes clearer if we quote from the original article:

> « *For example, as suggested by Fermi, the time-independent Schrödinger equation*

$$\Delta \psi(x,y,z) = (E-V)\psi(x,y,z)$$

---

[3] The copy of the letter can be obtained from the web site of the archives of the US government, under reference LAMS551. It is also reproduced in [Hurd, 1985]

[4] See for example a special issue of the Los Alamos Science, number 15 (1987) dedicated to Stan Ulam: several papers are personal memories and refer to the period. This review can be searched and downloaded from the Los Alamos library web site.

*could be studied as follows. Re-introduce time dependence by considering*

$$u(x,y,z,t) = \psi(x,y,z)e^{-Et}$$

*u will obey the equation*

$$\frac{\partial u}{\partial t} = \Delta u - Vu$$

*This last equation can be interpreted however as describing the behavior of a system of particles each of which performs a random walk, i. e. , diffuses isotropically and at the same time is subject to multiplication, which is determined by the function V. If the solution of the latter equation corresponds to a spatial mode multiplying exponentially in time, the examination of the spatial part will give the desired $\psi(x,y,z)$ - corresponding to the lowest eigenvalue E. »*

It is worth reformulating this argument more explicitely. Starting from the Schrödinger equation for the wave function of a system of N particles of mass m interacting through a potential *V(R)*,

$$i\hbar \frac{\partial \Psi(R,t)}{\partial t} = -\frac{\hbar^2}{2m}\nabla^2\Psi(R,t) + V(R)\Psi(R,t) \qquad (1.1)$$

In equation (1.1), *R* is a 3N-dimensional vector representing the coordinates of the *N* particles. One writes the (formal) solution as a linear combination of the generally unknown eigenfunctions of the Hamiltonian :

$$\Psi(R,t) = \sum_i c_i \phi_i(R) e^{-\frac{iE_i t}{\hbar}} \qquad (1.2)$$

where the sum is over all eigenfunctions of the Hamiltonian: the $c_i$ are the projections of the initial wave function on the eigenfunctions and the $E_i$ are the corresponding eigenvalues. By adding a constant to the Hamiltonian, one can always shift the eigenvalues by a constant, $V-E_0$, so that the the lowest eigenvalue is zero .For imaginary times, the sum is made of exponentially decaying functions except for the ground state, the state with the lowest eigenvalue (set to 0) which is therefore the solution for $t \to -i\infty$ . A change of variables $\tau = it/\hbar$ (and a shift in energy) changes the original Schrödinger to the following equation[5]

$$\frac{\partial \Psi}{\partial \tau} = D\nabla^2\Psi - (V-E_0)\Psi \qquad (1.3)$$

This equation has a form similar to what is known as a reaction-diffusion equation. The first term of the left hand-side is the diffusion term and can be simulated by a random walk of particles; the second term can be interpreted as a rate of appearance or disappearance of the « density" represented by the wave function (what Ulam names multiplication). To benefit from the idea of density and random walk, one has to consider a wave function which is and remains positive: this is always possible if one restricts oneself to study the ground state of systems of bosons or if one can neglect the quantum statistics. The case of fermions will be considered at a later stage. Both terms, the random walk term and the reactive, or branching, term, can be simulated on a computer. The invention of this probabilistic problem to solve the Schrödinger equation was the content of the idea contained in Ulam's article and credited to Enrico Fermi. We will come back to equation (1.3) later in the article when dealing with the Green Function method used to solve it.

---

[5] In the following, we will write *t* rather than tau for the imaginary time to simplify the notation.

The first application of the idea contained in Ulam's article was due to Donsker and Kac [Donsker, 1950]. Mark Kac developed the method involving a random walk so that it would eliminate the multiplicative (reaction) process: the possibility of the simplification of the numerical aspect was however limited to the test case of the linear harmonic oscillator. Donsker and Kac used random numbers stored on punched cards[6] on an IBM Electronic Calculating Punch , Type 604: their work necessitated several days of "computing" time to produce estimates of the first two eigenvalues of the quantum linear harmonic oscillator. With a slow convergence for the smallest eigenvalue and a second eigenvalue off by a factor 2: the computing requirements, even in the simple case of the harmonic oscillator, were way too high given the available resources at the time. The paper was published more as a technical report in a second-rank review and did not lead to further work before a long time had passed.

After the Donsker and Kac's article of 1950, one had to wait some fifteen years before another significant progress was achieved. The work for this next step was published in 1965 [McMillan, 1965], it was done by Bill McMillan who was a PhD student under the supervision of John Bardeen at the University of Illinois at Urbana-Champaign (UIUC). In the paper, David Pines and Leo Kadanoff , who both were at the Physics Department at UIUC, are thanked for inspiring discussions, Lars Onsager is also mentioned for providing a reference.

The system under study was He[4] at *T=0*. The model chosen for the Helium atoms is made of spinless bosons, interacting through a Lennard-Jones pair potential with parameters fitted from high-temperature experimental data. There were 32 particles in a box with periodic boundary conditions. The density was controlled by varying the size of the box, while the temperature was kept at zero degree, since the method was meant to determine the ground state wave function.

Rather than a direct simulation of equation (1.3) to find a stationary density leading to the ground state value, McMillan proposed to start from a reasonable trial wave function for the ground state followed by a procedure using random numbers to perform a variational calculus minimizing the energy and converging to the ground state. It is worth providing some more details on the procedure.

The trial wave function proposed is known as a « Jastrow » wave function:

$$R \Leftrightarrow (\vec{r}_1, \vec{r}_2 ... \vec{r}_N)$$
$$\psi^T(R) = \prod_{i<j} \exp\left[-\left(a_1 r_{ij}\right)^{a_2}\right] \quad (2.1)$$

---

[6] *A million random digits with 100,000 Normal Deviates* , Rand Corporation. Those numbers were obtained from measurements in a physical experiment. They were made available to the community on punched cards. Nowadays they are printed in a free ebook which can be dowloaded from the following web: https://www.rand.org/pubs/monograph_reports/MR1418.html. Interesting historical details are provided in the introduction.

In the previous equation, *R* is again being used for the ensemble of the *N* particle coordinates, $r_i$ (*i=1,...N*). The index *T* for the wave function indicates that it is a trial function, the starting point of the variational calculus. The two parameters $a_1$ and $a_2$ are the variational parameters, i.e. their value will be chosen under the criterion that they minimize the energy. The choice of this trial wave function was known to be a good starting point: It differs from the products of one-particle functions which is the usual starting point in quantum chemistry. This choice is possible since the focus is on the ground state for bosons: this permits to consider a wave function always positive, with no nodes as they would appear in excited states (or in Fermions systems, see section 4). Let us stress that postulating the Jastrow wave function as a starting point, is not so strong an approximation as it may appear: indeed, the procedure is to start from the most reasonable, and approximate form, and to converge to the exact ground state wave function. There are various reasons supporting the choice given in equation (2.1), for example that the function resembles the asymptotic solution of the two-body problem for small distances and rapidly becomes a constant at distances larger than interatomic spacing.

The work was done in the early 60s, at a time when it was already well widely accepted that Monte Carlo was an exact and efficient method to compute equilibrium properties in the classical statistical mechanics of liquids[7]. In particular the so-called Metropolis algorithm [Metropolis,1953] was known to be very efficient in sampling configuration space in statistical ensembles. This algorithm was in fact invented by Marshall Rosenbluth in 1953: it consists in a biased random walk in configuration space, with rules such that the generated trajectory samples the Boltzmann factor, $e^{-\beta E(R)}$ ( $with \beta = 1/k_B T$ ), of the canonical ensemble. A similar argument was then used by McMillan to sample the particle coordinates, *R,* with the wavefunction squared, being the probability, playing a similar role to the Boltzmann factor of the original Metropolis algorithm. Metropolis algorithm has proven to be one of the most important in computational sciences as it generates a set of configurations which are precisely the most contributing to the computed average values.

The actual sampling of the wave function is very similar to that used in classical ensembles. Particles are moved one by one to new positions by the addition of a random vector **x** (uniformly distributed in a cube of side D) to their new position. The new position is accepted with a probability

$$P = \min\left[1, \frac{|\psi(R_{new})|^2}{|\psi(R_{old})|^2}\right]$$

(2.2)

so that the set of the configurations generated during the biased random walk will sample the square of the wave function. The average value of any operator depending on positions is the arithmetic mean, along the walk, of its value with the positions generated along the walk.

$$\langle \hat{A} \rangle = \frac{\int dR A(R)|\Psi(R)|^2}{\int dR |\Psi(R)|^2} = \frac{1}{Nstep} \sum_{i=1}^{Nstep} A(R^{(i)}) \quad (2.3)$$

It is worth mentioning that, since the system is at *T=0*, there is no statistical ensemble sampling in addition to the wave function sampling: so that the average properties are readily obtained. For example, it is possible to express the energy in the form of a function of the positions and minimise the values of the parameters in the trial wave function. One can also compute in this way many static averages: the pair correlation function which is characteristic of the liquid or solid state of the system permit to determine the regions of phase stability as a function of the density.

---

[7] see for example [Battimelli,2020] or [Mareschal, 2018] and [Levesque,2018]

The results obtained in the work by McMillan were quite significant. The method which was developed was clearly opening a new way of dealing with many-body quantum systems and it allowed a direct comparison with experimental data ; in particular McMillan's work showed that the fluid phase of the chosen helium model remains stable within a range of reasonable (i.e. comparable to experimental ones) densities, while the solid phase appeared more stable at higher densities, corresponding in fact to higher pressures. The phase transition fluid to solid was however not reproduced by the simulation, given all limitations set by the very small system size considered: however phase coexistence was observed between the fluid and solid phases of Helium. The problem was studied at a later stage by Jean-Pierre Hansen, while on a sabbatical at Cornell University in the early seventies [Hansen, 1972].

It is also to be noted that this technique, known as Variational Monte Carlo (VMC), is still used nowadays with technical improvements over the basic technique imagined by Bill McMillan. Very unfortunately, Bill McMillan died at an early age, victim of a bicycle accident close to the campus of the University of Illinois where he was working.

# Malvin Kalos and the Green Function Monte Carlo.

The next progress in the MC method consisted in the development of the so-called Green Function Monte Carlo (GFMC) technique. This was achieved by Malvin Kalos in a series of papers, ranging from the early sixties till the late seventies : first focusing on small systems, 3 and 4-body nuclei [Kalos,1962], one helium atom [Kalos,1967], till a system of $N$ Lennard-Jones [Kalos,1970] and/or hard-sphere particle models [Kalos,1974a].

Going back to the case of our equation (1.3), its Green Function solution of the Schrödinger equation can be written as

$$\Psi(R',t) = \int dR\, G_{\Delta t}(R',R)\Psi(R,t-\Delta t) \qquad (3.1)$$

where the Green Function $G$ can be interpreted as the probability of moving from R to R' in a time interval $\Delta t$. It can be written in the position representation as

$$G_{\Delta t}(R \to R') = \langle R' | e^{-\Delta t\left[D\nabla^2 + (V-E_0)\right]} | R \rangle \qquad (3.2)$$

This is a formal way of writing the solution since we do not know what is actually the exponential of an operator in the position representation. The great merit of Malvin Kalos was to find an exact Monte Carlo way to sample $G$ even when it is unknown. The procedures to achieve this can be fairly complicated and rather than describing them, we discuss here a short-time approximation which leads to an explicit analytical form. If we let the time interval be small, we can then neglect the fact that the two operators in the exponential do not commute and write the following short time expression for the Green Function

$$G_{\Delta t \to 0}(R \to R') \cong \langle R'|e^{-\Delta t[D\nabla^2]}|R\rangle \exp\left(\Delta t\left[E_0 - \frac{V(R)+V(R')}{2}\right]\right) \quad (3.4)$$

or

$$G_{\Delta t \to 0}(R \to R') = w_D(R \to R').w_B$$

with

$$w_D(R \to R') = \frac{1}{(4\pi D\Delta t)^{3N/2}} \exp\left(\frac{(R-R')^2}{4D\Delta t}\right) \quad (3.5)$$

$$w_B = \exp\left(\Delta t\left[E_0 - \frac{V(R)+V(R')}{2}\right]\right)$$

The expression given in (3.5) becomes quite direct to understand: the first factor, with the D index, is a gaussian which governs the random walk part of the simulation to perform, while the second factor is the result of the branching process, giving a weight to configurations depending on the relative value of the local potential with respect to the reference energy. There are several ways to realise such a simulation: for example, considering an ensemble of configurations and letting them diffuse and then apply a branching mechanism, consisting in adding or removing members of the ensemble depending on the relative value of $w_B$, as indicated in equation (3.5). What clearly appears is that, within this short time approximation, successive applications of diffusion and branching, in an iterative way, will make any initial wave function converge to the ground state of the system with Hamiltonian *H*:

$$\Psi^{(n+1)} = \int dR\, G_{\Delta t \to 0}(R, R')\Psi^{(n)} \quad (3.6)$$

This short time version of the GFMC is, as a matter of fact, known as Diffusion Monte Carlo (DMC): it has the advantage of simplicity as compared to the original approach developed by Malvin Kalos. There are arguments concerning the accuracy of the method, as compared to the original approach. One very obvious limitation is the short-time limit, which introduces an approximation in the result. How important is the uncertainty introduced is arguable: arguments about this have been made and the discussion can become rather technical. The original method by Malvin Kalos did not use this short time expansion and he argued that it makes his original approach more exact. However, as one can imagine, each method has pros and cons but, in the end, both lead to very similar results (see, for example the discussion in [Cerperley,1983]). DMC has the important advantage of simplicity. More important is that, as such, the two algorithms, GFMC and DMC, are not very efficient: the origin can be traced in the branching factor. The large variation of the potential with respect to the reference energy makes it very difficult to converge without having a very large number of ensemble members. We will come back to this point in the next section, while discussing how adding importance sampling has dramatically improved convergence.

The basic ideas of the Green Function method were inspired from transport theory studies in radiation and nuclear physics. Mal Kalos made his thesis at the University of Illinois in the years ranging from 1949 till 1952, working on one of the few university computers available at the time.

From his thesis work, there is a paper published with J. Blatt [Blatt,1953] on nuclear physics. Kalos then moved to Cornell for a postdoc in the Department of Nuclear Physics, which was chaired by Hans Bethe at that time. He later joined a private company, *United Nuclear Corporation*, doing research in nuclear engineering from 1954 till 1965. While working in that company, he published several articles, among which can be distinguished a review written in collaboration with G. Goertzel[8] on Monte Carlo methods for the study of nuclear transport[Goertzel,1958]. The theory of transport of neutrons in reactors is based on an integral equation similar to the (linear) Boltzmann equation. It needs to be integrated numerically, and one of the possible techniques is based on a Green function method. There are several Kalos publications of that period which refer to an extension of those Green function's methods to the Schrödinger equation for a set of boson particles: first establishing an integral version of the Schrödinger equation, before formulating the solution with a Green Function and imagining a random walk procedure for the numerical implementation on the computer.

At some point, Kalos decided to leave the private company, *United Nuclear Corporation*, in which he was employed in a research and development position: he was experiencing the difficulty of addressing research problems of a more academic or fundamental nature. In 1965, he accepted a research contract at the Courant Institute of Mathematical Sciences : while the position he got was not really at his level of career development[9] (it was not an academic position), it offered him an unique opportunity for doing his own research. First, it was giving him the possibility to use the Courant Institute computer. This was the new Control Data (CDC) 6600, designed by Seymour Cray. This computer really made a difference at the time and it opened the way to high-performance computing: in the mid-sixties, it was certainly the best in the market for scientific computing. Kalos had always been knowledgeable about computers: he was easily programming in assembly language, and, having witnessed the evolution of both hardware and software from the very early years, he knew exactly what the computer was doing and how to optimize the running of any computer program. Besides, his role at the Courant Institute was providing him with an easy access to the CDC which he would use more or less freely during the weekends.

He was also in one of the best place for academic research, with all the New York scientific community around. He could rapidly establish connections leading to collaborations: in particular, he got connected with Loup Verlet who, with his group, had just started to work on the computational statistical mechanics of classical liquids, developing techniques for molecular dynamics studies of Lennard-Jones liquid and solid models [Levesque,2019]. Loup Verlet visited and collaborated with Joel Lebowitz who, either at Yeshiva University or later at Rutgers University, had become the compulsory meeting point for all new developments in statistical mechanics at that time. Kalos also kept contact with the Physics Department at Cornell, where he had been postdoc a few years earlier and where he shared an NSF research grant with Geoffrey Chester.

---

[8] In a book preface (*Monte Carlo Methods*, by Malvin H. Kalos and Paula A. Whitlock, J. Wiley, 1986, Gerald Goertzel was thanked for « *having introduced one of us to the mixed joys of Monte Carlo on primitive computers* ».

[9] Mal Kalos often points out in a joking manner that his career's bifurcations were very much like a random walk.

The GFMC had matured to the point of being able to provide the thermodynamics of a quantum model for a fluid or solid at T=0. The collaboration of Mal Kalos with Loup Verlet and Dominique Levesque led to a set of papers providing a method to develop the thermodynamics of quantum liquids. In [Kalos,1974b], a set of 256 particles modelling Helium 4 (a boson!) was being considered. Interactions were modelled as Lennard-Jones or hard spheres and , using the technique developed earlier,  properties were computed for solid and fluid phases (the driving parameter being the density). The papers can be considered as the equivalent quantum version to the classical theoretical description of the Lennard-Jones or hard-sphere models; the latter was achieved 15 years earlier and had a tremendous impact on classical statistical mechanics. This work was of course limited to zero temperature, however it made Kalos the world-expert on quantum Monte Carlo at the time.

The next step in the development of the Monte Carlo technique for quantum systems came from the sharing with Geoffrey Chester of the guidance of one of his PhD student, David Ceperley.

# The Fermion challenge

So far, we have been reporting on work done with bosons in their ground state. For bosons, as it has been stressed already, the ground-state wave function can be kept real and positive, making it easier to interpret the wave function as a density of particles and therefore to simulate the diffusion of those particles, together with the « multiplication » of ensemble members. The boson wave function remains identical under the permutation of any two particles.

$$f(1,2)=+f(2,1) \quad (4.1)$$

In the case of fermions, we have an antisymmetry property: the wave function changes sign under exchange of positions and spins of any two particles, so that it contains positive and negative regions in configuration space. The boundaries between regions of different signs, where it vanishes, are the nodes, or nodal boundaries (in a *3N* dimensional space, the boundaries are difficult to vizualize). While using products of one-particle function, it is possible to either symmetrize or anti-symmetrize the products. A general form of an antisymmetrical product of one-particle functions is given by what is called a Slater determinant (a determinant changes sign under permutation of two of its lines or columns, and vanishes whenever two of lines or columns are identical). In general, for *N* fermions, the wave function in the *3N* dimensional configuration space will have regions of positive and negative values and nodal surfaces to separate them.

Once Kalos had succeeded in finding a scheme to study bosons, the obvious question, was : "*what about fermions?*" . Geoffrey Chester, at Cornell, and Mal Kalos, at the Courant Institute, were sharing an NSF grant in a collaboration: they proposed the fermion challenge to a young graduate student who had started his thesis under the guidance of Chester in Cornell, David Ceperley. At that time David Ceperley had already started his PhD with a work on the classical one-component plasma [Ceperley,1977a]. He accepted to move to New York city, at the Courant Institute, to continue his work, learning MC techniques to describe neutron matter and extending these Monte Carlo techniques to fermion systems.

*During this period I worked in the Courant Institute at New York University (NYU) in Kalos's group. Mal Kalos has been deeply influential throughout my career since that time. During my 3 years at NYU Kalos taught me all about Monte Carlo and computation as no one else could. He was an expert with many years experience in Monte Carlo, in finding very clever and non-standard sampling methods, in assembly language programming and was the world expert on quantum Monte Carlo. I attended a semester long class he taught on Monte Carlo and learned many topics I have later used such as correlated sampling, importance sampling, statistics; topics I later incorporated into classes I later taught. The book, Monte Carlo Method [Kalos,1986], is a readable summary of the course material.*

*The Courant Institute at that time was a very inspiring collection of mathematicians, computer scientists and physicists. I had access to the supercomputer of the day, a CDC 6600. The input to it was through punch cards. Jobs would be run through a batch system and output was on a line printer. Kalos would come in on Saturdays and run the computer by himself along with his graduate students and postdocs. Chester would visit for a day or so each month to check on and advise me.*

*Mal and Geoffrey together showed me how valuable collaborations could be. Also associated with the group were Paula Whitlock, Mikkilineni Rao, Jerry Percus, Joel Lebowitz and Jules Moskowitz. We had visits by Dominique Levesque, Oliver Penrose, Kurt Binder and Luciano Reatto from Europe. Daily life often included lunch at inexpensive restaurants in Greenwich Village or in Chinatown and every afternoon, "tea" on the top floor of the Courant Institute with the Institute staff.*

The work on fermion systems was published in 1977 and the article is signed by Ceperley, Chester and Kalos, we refer to it as the CCK paper[Ceperely,1977b] : it is the first MC simulation of a many-fermions system. The starting point is a trial wave function which generalises the boson case in a direct way:

$$\psi^T(R) = \psi^T_{bos}(R) |u_n(i)|_{Slater} \qquad (4.2)$$

where the $u_n(i)$ are one-particle wave functions, i. e. plane waves for a delocalised system (an electron gas or a liquid) and Gaussian orbitals for a solid with atoms centred on lattice nodes, for example. The index $i$ refer to particles, whereas the index $n$ refers to wave-function numbering[10]. Those wave functions are arranged in the Slater determinant form in order to contain all the antisymmetry requirement of the total trial wave function.

There are two obvious questions to proceed: how costly is it to compute the determinant at each time step of the random walk? , and second : how to treat the jumps close to the nodal boundaries, i. e. when the random walk moves a particle from a region where the wf is positive (resp. negative) to a region where it is negative (positive)?

The calculation of the full determinant at each time step is in principle very costly and, instead, the procedure was set to move one particle at a time and then, not recalculate but rather updating the determinant using formulas of linear algebra: since you changed only one particle position, you have only modified one line, or one column, of the determinant, so that the new value can be obtained by multiplying by a quantity needing in its computation a factor N less operations than the full determinant. David Ceperley mentions that he found the inspiration for doing this while browsing through discount books at a bookstore near NYU and finding by chance (!) a book with the useful formulas [Gastinel, 1970]. Updating the determinant after one particle's move rather than re-computing the full determinant saves a factor $N$ (number of particles) in the computation time. Without this Bookstore's inspiration, the computation would not have been possible at that time!

The second question is more difficult to handle and a complete solution is still missing at this time. The Slater determinant fixes boundaries between regions with positive and negative values of the wave function. The random walk procedure used was similar to the McMillan's variational method, inspired from the Metropolis algorithm: it is therefore independent of those nodal boundaries and it allows motion of configurations across the original nodal boundaries. How much sensitive will be the result on the precise location of the boundaries

---

[10] The index n should refer to both the spatial and the spin wave function; in order to make reading easier, we only speak about spatial dependence.

A good solution to this problem will be found later: it will be to require that the full wave function keep the same nodal boundaries as the trial wave function. This is an approximation (known as the *fixed-node* approximation) and it is variational in nature. The energy computed will be an upper bound for the true energy of the system to simulate. Several other methods have been devised to treat fermions, all appear to have some drawbacks. The fixed node approximation is simple, stable and give accurate results in in the many-fermion problem.

The CCK paper appears retrospectively as a real breakthrough, providing the basics of a numerical technique leading to extend QMC results for dense fermion systems, an opportunity which was later used more and more with the availability of cheaper computing[11]. In 1977, the technique was applied to $He^3$ and some models of neutron matter, with a critical discussion of previous approximate results. One of the clear merits of the paper is to dismiss the widespread belief that it is not computationally feasible to sample a fully antisymmetric function for a reasonably large number of particles (of the order of 100). An important review paper was then published summarising the state of the field in book dedicated to Monte Carlo methods and edited by Kurt Binder [Cerpeley, 1979].

In 1985, Mal Kalos became professor at the Courant Institute of Mathematical Sciences, and was appointed Director of the Ultracomputer Research Laboratory. In 1989, he moved to Cornell where he was offered a professorship in the Physics Department, as well as the Directorship of the Cornell Center for Theory and Simulation in Science and Engineering. Mal Kalos kept using the Green Function MC method, however the method was not much used by the community as it appeared more complicated than Diffusion MC or Variational MC.

# Meeting Berni Alder

The calculation of the ground state energy of the electron gas [Ceperley,1980] is the most celebrated exact result calculated by Monte Carlo: this computation has been achieved a few years later in a paper resulting from the collaboration of David Ceperley and the founding father of molecular dynamics, Berni Alder. At the occasion of the $50^{th}$ anniversary of the Physical Review Letters, this paper was ranked number 3 in the list of the ten most cited and important papers of the journal[12]. The paper reports on the computation of the exact ground state of the electron gas, i. e. $N$ electrons, interacting through their Coulomb repulsion in a box with periodic boundary conditions. This calculation has been used as one of the basic building blocks of the first-principle molecular dynamics technique, the latter being more and more developed in the following years and becoming widely used in materials science and statistical mechanics. As a matter of fact, this calculation used technique which was developed after the CCK paper by David Ceperley and which will be shortly described below.

---

[11] A 128 fermions systems necessitated 5 seconds on the CDC 6600 for a single pass (each particle is moved once on average). This appeared costly in 1977. Petaflops computers, which are common nowadays, are 10 orders of magnitude faster than the 6600!

[12] *APS News*, June 2003 (Volume 12, 6)

After the completion of his thesis, David Ceperley spent a postdoctoral year in Orsay, within the Verlet group and another year in a postdoc position offered by Joel Lebowitz in New York. The year spent in Orsay (1977) did not lead to a collaboration with Loup Verlet who had, for personal reasons, left the field. Loup Verlet [Levesque, 2019] had been the founder of a school of computational physics in Orsay: this school developed successfully in the following years, however Verlet himself did not contribute personally any more to research. The postdoc spent in France gave Ceperley time to reformulate Kalos' GF technique in terms of the simpler diffusion Monte Carlo method.

The starting point of Ceperley's reformulation of the Monte Carlo technique is the trial wave function defined by equation (4.2). The function obeying the Schrödinger equation will be denoted $\Phi(R,t)$; we also let their product be denoted

$$f(R,t) = \psi_T(R)\Phi(R,t)$$
where  (5.1)
$$\frac{\partial \Phi(R,t)}{\partial t} = -H\Phi(R,t)$$

The trial wave function is not an eigenfunction of the hamiltonian and we can denote the action of H on it by another function of the configuration space $R$, namely:

$$H\psi_T(R) = E_T(R)\psi_T(R) \quad (5.2)$$

This defines a local energy which varies in configuration space: it would reduce to a constant (the ground state energy!) if the trial function would approach the exact ground state wavefunction. The equation obeyed by the product between the trial wave function and the solution in time of the Schrôdinger equation can then be derived. It reads:

$$\frac{\partial f(R,t)}{\partial t} = D\left[\nabla^2 f - \vec{\nabla}\cdot\left(f\vec{\nabla}\ln|\psi_T|^2\right)\right] - \left[\frac{1}{\psi_T}H\psi_T - E_0\right]f(R,t)$$

(5.3)

This equation for the product *f(R,t)* is the equation being really simulated in DMC. It is similar to the equation for the wave function, equation (1.3), with two crucial differences. First, in addition to the diffusion term, there is a drift term which accelerate in the direction where the function *f(R,t)* is important. Second, the branching part is now determined by the difference between the local energy, defined in equation (5.2) and the reference energy.

We could repeat the steps taken to solve the Schödinger equation of section 3, leading to the same expressions for the jump and branching probabilities, eq.(3.4) and (3.5), with the substitutions

$$R - R' \rightarrow R - R' - D\Delta t F(R)$$

*with* (5.4)

$$F(R) = \nabla \ln |\psi_T|^2$$

and

$$V(R) \rightarrow E_T(R) \qquad (5.5)$$

The procedure is very similar to the one described in the previous section: there is a drift in addition to the random walk, driving the system in important regions of configuration space. Besides the branching is now determined by the local energy which is almost a constant[13], unlike the potential *V(R)* of course. Both the local energy and the drift force have to be computed at each timestep, but the computation is done in the important regions of configuration space. It is another example of *importance sampling*. The simulation is much more efficient and more accurate than the original one[14].

Another property of equation (5.3) is that the nodal boundaries remain fixed: whenever we approach a nodal boundary where the trial function vanishes, the drift force becomes infinite, pointing in the direction away from the boundary. True nodal surfaces are expected to be different from those obtained by products of one-particle wf, probably not very different as it can be argued, from the non-interacting case. Later on, David Ceperley considered releasing the nodal boundary surfaces (*released node* algorithm) to improve on that approximation: particles would be allowed to cross the nodal surfaces which would not be changed during the computation. In doing that, one avoids potential non-ergodicity of the walks: with particles trapped in regions, one would not be able to sample the entire phase space. Later on also, David Ceperley proved a so-called « tiling theorem »[Ceperley,1991] which states that all nodal regions (at least within a mean-field theory) have similar properties so that it is not necessary to allow crossing the nodes to have an ergodic sampling. Several important results have been reported since the CCK paper, without really fully solving the issue.

To simulate Brownian motion, a technique had been developed in the statistical physics community, know as *Brownian dynamics*. To extend the method to the equation for *f(R,t)*, one needs to generalise Brownian dynamics by adding the branching process and the density would reach, asymptotically in time, the stationary solution: the latter is named the mixed distribution and it is the product of the trial wave function times the exact ground state wave function. The ground state energy is obtained as the average of the local energy over the mixed distribution. As emphasised in a paper by Ceperley and Alder [Ceperley,1981], « *the computation of the properties of the many-*

---

[13] It would be a constant if the wave function would be the exact ground state wave function.

[14] It is worth mentioning that equation (5.3) bears some similarity with the Smoluchowski equation in chemical physics: the Smoluchowski equation has no branching term and it describes Brownian particles undergoing an acceleration due to an external force. It is also simulated by a stochastic method known as Brownian Dynamics. The branching term is essential in the present context as it describes all the interaction part of the N-body problem. Moreover, it is also the term which allows to introduce importance sampling.

*body wave function has thus been reduced to finding an accurate though simple, trial wave function and a sufficiently powerful computer, as these are the two factors which limit the accuracy of the method.* »

In the formulation given above for DMC, use is being made of *importance sampling*. This method is crucial for the convergence of Monte Carlo's calculation of multidimensional integrals. Instead of random sampling of the integrand - most of the points chosen would contribute nothing to the integral -, points are chosen according a distribution as close as possible to the real distribution of the integrand. In the present case, one starts from the trial function , obtained by VMC, and one concentrate the computational effort in the region where this trial function is important.

One should also mention the work by Jim Anderson who developed [Anderson, 1976], at around the same time, a random walk simulation method to solve the Schrödinger equation in chemical contexts. This work was more intuitive than Kalos' and Ceperley's approach and did not use importance sampling at its beginning. In 1979, Anderson and Ceperley collaborated on chemical applications of DMC. This resulted in converting Anderson to the usage of importance sampling[Anderson, 1980]: later, Jim Anderson systematically applied the method to various chemical applications of increasing complexity[Anderson, 2002].

This elegant formulation of the DMC method came step by step as the career of David Ceperley converged to California. While in France (1976-77), David met Joel Lebowitz who was spending a sabbatical at the Institut des Hautes Etudes Scientifiques, in Bures-sur-Yvette, near Paris. This collaboration led to a postdoc position being offered to David Ceperley while returning to New York in 1977. It gave David Ceperley an opportunity to work on polymer physics and on the computer modelling of their dynamics: the problem necessitated to adapt the existing (classical) simulation methods to the case where the convergence could be very slow, due to the long timescales of polymer relaxation. The work done at this occasion has proved to be very useful at several later occasions, in particular within the path-integral MC method which has much in common with polymer physics: this will be described in QMC2.

While in France, David Ceperley met Berni Alder at a summer school held in september 1977 in Aleria, on the island of Corsica. Berni Alder, at that time was known as the inventor of the Molecular Dynamics method [Alder, 1958] which he had successfully applied to various classical model systems of statistical physics. He was very then much interested in extending his Molecular Dynamics methods to quantum systems: Berni was impressed by David Ceperley and he understood the potential developments of his thesis achievements, as he said in an interview [Battimelli, 2018]. According to David Ceperley: « *I got to know Berni Alder quite well during this school and it was partially through this contact that I obtained a job offer in Berkeley a year later* ». This job offer was made in 1978 by the newly founded National Resource of Computational Chemistry (NRCC), within the Lawrence Berkeley National Laboratory in California: the centre was funded by NSF in order to develop and maintain software for computational chemistry. It was successful and developed a few popular codes (at least among the theoretical chemists), however NSF did not want to continue funding after three years and the centre was closed in 1981. David Ceperley then joined the Livermore National Laboratory in the group of Berni Alder with whom he had started collaborating since 1978. He stayed in Livermore a few years, until he accepted a Faculty position in 1987 at the University of Illinois at Urbana-Champaign.

At the beginning of this NRCC period, while being located in Berkeley, Ceperley started to collaborate with Berni Alder. However, he did not get immediately the necessary security clearance in order to be physically in Livermore. Access to the Livermore computers, discussion of the results then took place with the help of Berni Alder's collaborator, Mary Ann Mansigh[15]. Every morning, Mary Ann was calling David Ceperley by phone and they were discussing the results obtained during the previous day (and night) and what to do next. It is only after more than a year of procedure that the security clearance was granted.

The Berkeley-Livermore period was most productive for David Ceperley. Of course, the availability of the best computers of the time (at least in Livermore: a CDC 7600 and Cray-1 in 1978; a Cray-2 and cyber 205 later) was important[16]. But probably more important was the influence of Berni Alder: « *After Kalos, Berni was my most important mentor. He taught me the importance of attacking the big, hard problems and of making definitive calculations* ».

The first big problem has been the calculation of the electron gas which we have already alluded to, « *the calculation that made me famous* », as David Ceperley puts it. For Berni Alder, this was like a toy model, used to show that one could do an exact calculation on a N-body quantum system, just as Molecular Dynamics could do on classical N-particle models. The real challenge, according to Alder, would be the computation of hydrogen where one would be able to confront the experimental measurements, a material which was, and still is, strategic in many respects.

The quantum uniform electron gas at zero temperature, known as jellium, is a model for solids where the positive charges (the atomic nuclei and the core electrons) are assumed to be a uniform background in space, with no other role than realising overall electro-neutrality. This model is often used to derive the behaviour of metals. The driving parameter of the model is the electronic density. At high density the system is known to behave as an electron Fermi gas, as the kinetic energy of the electrons dominate the interaction energy. The electrons are delocalised and the balance of spins up and down leads to a paramagnetic state. Decreasing the electron density, there is first a transition to a ferromagnetic Fermi fluid as interaction energy starts being more important: it becomes energetically more favourable to have an imbalance between up and down spins. At sufficiently low density, one has the formation of a crystal, the so-called Wigner crystal where single electron orbitals become of Gaussian form and are centered on lattice sites.

The 1980 DMC calculation of Ceperley and Alder [Ceperley,1980] was showing that QMC provides the most accurate quantitative approach to determine the phase diagram of jellium. The paper reports on the transition from a paramagnetic to ferromagnetic phase as well as the transition to a crystal structure as density is being lowered. Subsequent QMC calculations have refined the original calculations: the transition to ferromagnetism is believed to be second order [Zong,2002]. The two-dimensional Fermi fluid has been studied as well [Tanatar, 1989] leading to a somewhat different phase diagram than in three dimensions.

---

[15] An interview of Mary Ann Mansigh by Daan Frenkel can be viewed among the Cecam lectures, on the Cecam web site (https://www.cecam.org/cecam-lectures/)

[16] Besides, the Alder group had the privilege, through a friendly and totally unofficial agreement with the computer centre, to use freely the idle time. Whenever the large hydrodynamic codes of the bomb designers would crash, and that was happening quite frequently, the computer would restart a QMC code from the last written configuration.

In the 1980 calculation, the number of electrons considered was between 38 and 250, with an average number of ensemble members of around 100. This could allow an examination of number dependence in the results and an extrapolation to an infinite system. The authors mention that they could also estimate the uncertainty due to the timestep which proved to be quite small. They used the fixed node algorithm, with absorbing barriers as boundary conditions: any electron position which would cross a nodal boundary during the simulation would lead to remove the electron from the simulation. They also mention considering another procedure which they call nodal relaxation: one releases the nodes so as to allow the energy to decrease: when a random walk strays across a nodal boundary, it is not terminated but the sign of its contribution to the average is reversed. Some contributions are counted positive, some are negative and it is the difference population which converges to the antisymmetric eigenfunction. The important point is that the positive and negative parts have similar behaviour, that they do not grow too quickly. This is what happened in the electron gas. According to Berni Alder: »But, if you go to other systems, like chemical systems, where the difference between the Fermi and the boson energy is very large, you can no longer accurately project out the difference. » The numerical instability , in that case, is known as the fermion sign problem and a general solution is still missing.

The important point -which had a tremendous impact on the community- was that the QMC calculation was providing a way to accurately interpolate the electron gas properties into the important range of electron densities for atoms, molecules and metals: "*it did produce a correlation energy for spin-polarized electron gas with a small estimated error and could determine the non-local corrections to the local spin density approximation in real systems* » [Vosko, 1980]. It turned out that this electron gas calculation became a standard, the most referenced paper in the field[17]. This came about because David Ceperley shared his preliminary results with several groups that were working to improve the density functional theory (DFT) exchange and correlation energy. Quoting David Ceperley: « *Two publications [Vosko,1980] [Perdew,1981]gave my QMC calculation a great deal of publicity. Partly because of Ceperley Alder correlation energy, the whole DFT enterprise took off in the 1980s. There are a lot of problems with DFT but the basic ingredient, the exchange and correlation energy of the uniform electron gas became a standard of that time. DFT users did not have to consider it as an unknown variable. Twenty years later, when the community moved beyond standard LDA (Local Density Approximation), a tower of Babel of energy functionals have re-emerged* ».

After this success, with Berni Alder urging, they started the work on hydrogen[18]. As explained by Ceperley: « *An electron gas is not directly realized in any material, it's an idealized model, while hydrogen is a real material. With the hydrogen calculation we wanted to address experimental predictions, not just compare with theory. Our hydrogen calculation was the first many-electron*

---

[17] It ranked number three in the list of most cited papers published by PRL, a ranking made in 2002 by the APS-while PRL started in 1958. Number 1 was S. Weinberg 1967 paper on the electro-weak interactions, number 2 was M. K. Wu 1987 letter on high Tc superconductors, while the 4th was Binnig's paper on Atomic Force Microscopy and number 5 was 1985 Car-Parrinello's seminal paper on Density Functional Theory.

[18] It is worth mentioning that David Ceperley never wrote the long PR paper which was originally planned on electron gas: David Ceperley  confesses that it was harder than previously thought to improve on it. And also as Berni Alder puts it: "*after I finish a problem, I like to move on*".

*calculation of a material to lead to important predictions. Berni's suggestion was excellent, 40 years later there are plenty of unsolved problems both theoretically and experimentally for dense hydrogen. It is a very active field* ».

The importance of hydrogen is pretty obvious: it is the material which is the fuel of the fusion reaction, whether in controlled or uncontrolled way, both ways being equally strategic in Livermore. It is also the most abundant component in stars in the universe, where it is often undergoing very high pressures, like 10 million times the atmospheric pressure. Those pressures cannot be realised experimentally, whether in shock wave experiments or in diamond anvil cells. A theoretical calculation, where all forces between constituents are of fundamental nature, was highly desirable. There had been many theoretical predictions on hydrogen: for example, in the early thirties, Wigner and Huntington predicted [Wigner, 1934] that hydrogen would become metallic at extremely high pressures (one fourth of a million times the atmospheric pressure)!

The hydrogen computation [Ceperley, 1981] was hard and more difficult than the electron gas: there were both electrons and protons (but no core electrons), hydrogen could be in molecular or atomic form, one would have to consider a metallic phase with delocalised electrons and compare the relative stability of those various phases. But the important thing was that the VMC/DMC methods could be used to tackle the problem. According to Ceperley: « *Our calculations starting in 1980 had electrons and protons with the protons on a BCC lattice, and the electrons in a delocalized metallic state. We would then calculate the energy, change the crystal structure to determine which structure was most stable and vary the density. After we did metallic hydrogen, we then did molecular hydrogen so we could see the phase change between metallic and molecular hydrogen. We predicted the transition at around 3 Mbars. That might be right since experimentalists haven't quite gotten achieved those pressures reliably. But at the time we only considered very simple crystal structures, simple trial wave functions, and very crude finite-size corrections so our precise predictions are no longer to be believed.* »

The only approximations of those calculations were the finiteness of the systems considered ( a few hundred atoms), the use of fixed-node approximations at times and the finite length of MC runs. The most important paper on this subject was published in 1987 [Ceperley, 1987]. Since these early calculations, with increased computing power, and new methods being developed, most of the phase diagram of hydrogen has been computed and successfully compared to experiments (see [Pierleoni,2006]). Both protons and electrons have been treated quantum mechanically, going therefore beyond the Born-Oppenheimer approximation. The latter (electrons relax instantaneously to proton moves) is made in most theoretical descriptions. Quoting David Ceperley again: « *Since protons are 1836 heavier than electrons they move that much more slowly, so, just as in molecular dynamics, you have a two-time scale problem: the time step is determined by the electron-electron and the electron-proton collision times. The electrons have to move thousands of time steps before the protons move very far. You may need to have millions of time steps before the proton's distribution will converge. I also had to develop trial wave functions that included not only electron-electron and electron-proton terms and electronic orbitals, but also proton-proton correlations and protonic orbitals. It is with the protonic orbitals that the physical phase is determined.* »

# Conclusions

The first, obvious, point which emerges from that story is the long time needed to have a brilliant idea, suggested in 1949 Ulam's paper, transformed into a working and useful algorithm in 1980. David Ceperley and Berni Alder could benefit from ample computing time on the fastest machines of the day in order to achieve this. Equally, or even more important, was the maturation on the algorithmic part: David Ceperley extended Kalos'method to fermions, he also made the algorithm more direct and more clear, in such a way that both importance sampling and behaviour at wavefunction's nodes could be understood and mastered in a profitable way. The Ceperley-Alder collaboration came at the right moment, with the right people at the right place!

The 1980 computation of the electron gas came as a real breakthrough in electronic structure computations. It made possible the development of several methods of molecular simulations based on first principles: instead of using phenomenological interaction potential between atoms, these methods, developed from the electronic density functionnals, could base their model on the number of protons and electrons and use forces of fundamental nature. The correlation energy of the electron gas, provided by the QMC computation, was a key ingredient. This illustrates also how methodological progress in one field can impact other fields in ways that were not expected at the beginning.

This family of QMC methods, known as projection Monte Carlo, have continued their development, enlarging their modelling capacity as new methods are being developed and computing power becomes cheaper and approaches the exa-flop regime. Recent reviews allow to check on those developments: see [Ferrario, 2006] for example. Another family of QMC , based on path integrals and dealing with quantum many-body systems at finite temperature, have also been developed in the 1980s. It will be reported in a companion article.

# Acknowledgements

The idea to undertake this study was a suggestion by Giovanni Ciccotti whom I wish to thank for a constant support, a generous availability and many useful suggestions. I wish also to thank David Ceperley for his hospitality during my visit to him in Illinois and for later correspondance. Stimulating discussions with Benoit Roux are gratefully acknowledged.
David Ceperley's interview was made possible through financial support provided by the Neubauer's Collegium of the University of Chicago.

During the completion of this article came the sad news of the death of Berni Julian Alder the day preceding his 95th anniversary. I wish to acknowledge here his constant, generous and enthusiastic availability for sharing his memories as well as his views on the field he has contributed to create.